\documentclass[preprint,aps]{revtex4}

\begin{document}

\title{Proposed New Test of Spin Effects in General Relativity}

\author{R. F. O'Connell{\footnote{ E-mail: rfoc@phys.lsu.edu}}}

\affiliation{Department of Physics and Astronomy, Louisiana State
University, Baton Rouge, Louisiana  70803-4001}

\date{\today}

\begin{abstract} The recent discovery of a double-pulsar PSR J0737-3039A/B
provides an opportunity of unequivocally observing, for the first time,
spin effects in general relativity.  Existing efforts involve detection of
the precession of the spinning body itself. However, for a close binary
system, spin effects on the orbit may also be discernable.  Not only do
they add to the advance of the periastron (by an amount which is small
compared to the conventional contribution) but they also give rise to a
precession of the orbit about the spin direction.  The measurement of such
an effect would also give information on the moment of inertia of pulsars.
\\
\\
\\
\\
\\
\\
\\
\\
\noindent PACS number(s): 04.20.-q, 04.25.nX, 04.80.-y
\end{abstract}


\maketitle

The Hulse-Taylor binary pulsar PSR 1913+16 has proved to be a fascinating
laboratory-in-the-sky for the investigation of general relativistic
effects.  However, the recent discovery of a close double-pulsar binary
system \cite{burgay,lyne} (with an orbital period about 3 times smaller
and a pulsar period also smaller for the larger mass pulsar) promises an
opportunity for even more exciting discoveries.  Here, we consider spin
effects.

Attempts to measure gravitational effects due to spin in the laboratory
are futile \cite{oconnell1}.  The proposal of Schiff to
investigate such effects by measuring the precession of a small gyroscope
in earth orbit is the basis of the successfully
launched Gravity Probe B experiment \cite{reichhardt}.   Apart from
technological and scientific challenges \cite{oconnell2}, the largest
relativistic precession rates involved are less than about seven seconds
of arc per year.  On the other hand, with the discovery of PSR
1913+16 in 1975, it was immediately clear that much larger spin precession
effects come into play.  However, it was also obvious that the results of
Schiff were no longer applicable since we are now dealing with a 2-body
system.  The correct required 2-body spin precession result was provided
by Barker and the present author \cite{barker1,barker2,barker3} from which
it became apparent that the spin direction precesses about the orbital
angular momentum direction at a rate of
$1^{\circ}.213yr^{-1}$ i.e. a factor of about $6 \times 10^{5}$ larger
than for the earth gyroscope.  However, while efforts to measure this
precession have proved difficult,
\cite{kramer,weisberg} there is hope that observations on the new
double-pulsar will prove more fruitful because the calculated
\cite{barker1,barker3} spin precession rates are even larger \cite{lyne},
viz.
$4^{\circ}.8yr^{-1}$ for
$A$ and $5^{\circ}.1yr^{-1}$ for $B$.

We now point out that correspondingly larger numbers arise for spin
effects on the orbital motion which might prove easier to observe.
Observable implications of such effects have been explored for a
variety of astrophysical binary systems but no definitive results have
emerged \cite{barker4,damour,kopeikin,morsink,wex,iorio}.  In
particular, it would appear that the best possibility might be the
effect on the gravitational radiation waveform in coalescing binary
systems of compact objects \cite{kidder} but such hopes rest on the
detection of gravitational radiation.

On the other hand, the double-pulsar system presents the possibility of
a clean test \cite{burgay,lyne}.  The greatest effect on the orbital motion
in the double-pulsar system is the periastron precession, amounting to
$16^{\circ}.90yr^{-1}$.  More important is to note that, in addition to
these precessions about the orbital angular momentum direction, there is
also a precession of the angular momentum of the orbit itself about the
spin directions \cite{barker1,barker3} which only occurs due to spin
effects.  This is a reflection of the fact that the total angular momentum
of the whole system remains constant (as shown explicitly in
\cite{barker1}) so that a precession of the spins implies a precession of
the total orbital angular momentum (both effects resulting from spin-orbit
interactions and, to a lesser degree spin-spin interactions).  Explicit
results have been written down for these quantities,
\cite{barker1,barker3} which we now use to calculate the contribution from
the fast 23-ms pulsar (with a mass $m_{1}$, say) to this precession
(noting that the contribution of the 2.8 -sec pulsar, with mass $m_{2}$,
may be obtained from our general formulae by the simple replacement
$m_{1}\leftrightarrow m_{2}$).

Using the notation of \cite{barker1} and \cite{barker3}, the secular
result for the rate of precession of the orbital angular momentum
$\vec{L}$ due to the spin $\vec{S}^{(1)}$ of $m_{1}$ may be written as

\begin{equation}
\frac{d\vec{L}}{dt}=A^{*(1)}\left(\vec{n}^{(1)}\times\vec{L}\right),
\label{ntse1}
\end{equation} where

\begin{equation}
A^{*(1)}=\frac{GS^{(1)}\left(4+3m_{2}/m_{1}\right)}{2c^{2}a^{3}\left(1-e^{2}\right)^{3/2}}
\label{ntse2}
\end{equation} and where $\vec{n}^{(1)}$ is a unit vector in the direction
of $\vec{S}^{(1)}$, $a$ is the semi-major axis and $e$ is the
eccentricity. Writing

\begin{equation} S^{(1)}=I^{(1)}\omega^{(1)}=m_{1}k^{2}_{1}\omega^{(1)},
\label{ntse3}
\end{equation} where $I^{(1)}$, $\omega^{(1)}$ and $k_{1}$ are the moment
of inertia, the angular rotational velocity and the radius of gyration,
respectively, of $m_{1}$, we obtain

\begin{equation}
A^{*(1)}=\frac{G\left(4m_{1}+3m_{2}\right)\omega^{(1)}}{2c^{2}a\left(1-e^{2}\right)^{3/2}}\left(\frac{k_{1}}
{a}\right)^{2}. \label{ntse4}
\end{equation} Since, from \cite{burgay} and \cite{lyne}, $a=8.79\times
10^{5}km$ (which is obtained from the 2-body Kepler formula
\cite{barker1,barker2} and the measured period of orbital revolution
\cite{burgay,lyne}) and, for a neutron star, \cite{bignami} $r_{1}\approx
15km$ we see that
$\left(k_{1}/a\right)^{2}\approx 2.91\times 10^{-10}$.  Also
\cite{burgay,lyne}
$\omega^{(1)}=2.768\times 10^{2}$ rad/sec and
$\left(1-e^{2}\right)^{-3/2}=1.0117$.  Thus

\begin{equation} A^{*(1)}\approx 4.06^{\prime\prime}/yr . \label{ntse5}
\end{equation} This result for the rate of precession of the angular
momentum of the orbit about the pulsar spin direction is clearly
measurable over a sufficient period of time (and we note that it is more
than $10^{3}$ times larger than the ms-arc precession rates desired of the
gyroscope experiment) as it is reflected in a corresponding change in the
mass function which depends on observable quantities. Furthermore, it
could eventually provide a way to obtain accurate information of the
moment of inertia of neutron stars, particularly if both spin precession
and orbital precession are observed.  We note that there is also a
contribution from the spin of pulsar
$B$ to the precession of
$\vec{L}$ but since its pulse period is 122 times larger than that of
pulsar $A$, its contribution will be correspondingly smaller.  Finally, we
note that contributions of the same order of magnitude are present due to
spin contributions to the perihelion precession since the angular momentum
and Runge-Lenz vectors precess at the same rate \cite{barker1,barker3}
but, in addition, there is also a comparable contribution from
second-order spin-independent post-Newtonian effects \cite{damour,wex2}.
However, we feel that a measurement of the precession of the orbital
angular momentum is a cleaner test since any such precession is a definite
signature of spin effects.  The corresponding results for other
gravitational theories are given in \cite{barker5}.

\acknowledgments I thank Dr. M. Kramer for a very helpful exchange
concerning the work in Refs. \cite{burgay,lyne} and for drawing my
attention to Ref. \cite{wex2}.

\end{document}